\documentstyle[11pt,aaspp]{article}
\def\dl{\partial} 
\def\px{\phi(x)}
\def\be{\begin{equation}}
\def\ee{\end{equation}}
\def\bea{\begin{eqnarray}}
\def\eea{\end{eqnarray}}
\def\xio{\xi^{(1)}(x)}
\def\xioy{\xi^{(1)}(y)}
\def\xiot{\xi^{(1)}(x,t)}
\def\xit{\xi^{(2)}(x)}
\def\xitt{\xi^{(2)}(x,t)}
\def\nf{\nabla^4}
\def\n2{\nabla^2} 
\def\de{\delta_{\mu \nu}}
\def\xmn{\frac{x_{\mu} x_{\nu}}{x^2}}
\def\iox{\int_0^x \xioy}
\def\ioo{\int_0^{x_1} \xioy}
\def\ioi{\int_0^\infty \xioy}
\def\x0g{x_o^{\gamma}}
\def\lb{\left(}
\def\rb{\right)}
\def\Ca{\ioi y dy}
\def\Cb{\ioi y^2 dy}
\def\Cc{\ioi y^4 dy}
\def\ga{\gamma}
\def\np0{\nabla^2 \phi(0)}
\begin{document}
\title{Self-similarity and the pair velocity dispersion.}
\author{Somnath Bharadwaj \\ Mehta Research Institute, 10  Kasturba Gandhi
Marg \\ Allahabad 211 002, India \\ e-mail: somnath@mri.ernet.in }
\authoraddr{Mehta Research Institute, 10 Kasturba Gandhi Marg, Allahabad
211 002,India \\ somnath@mri.ernet.in}
\begin{abstract}
 We have considered linear two point correlations of the
form $\frac{1} {x^{\gamma}}$ which are known to have a self-similar
behaviour in a $\Omega=1$ universe. We investigate under what conditions
the non-linear corrections, calculated using the Zel'dovich approximation,
have the same self-similar behaviour.  We find that the scaling properties
of the non-linear corrections are decided by the spatial behaviour of the
linear pair velocity dispersion and it is only for the cases where this
quantity keeps on increasing as a power law (i.e. for $\ga<2$) do the
non-linear corrections have the same self-similar behaviour as the linear
correlations. For $(\ga > 2)$ we find that the pair velocity dispersion
reaches a constant value and the self-similarity is broken by the
non-linear corrections. We find that the scaling properties calculated
using the Zel'dovich approximation are very similar to those obtained at
the lowest order of non-linearity in gravitational dynamics and we propose
that the scaling properties of the non-linear corrections in perturbative
gravitational dynamics also are decided by the 
 spatial behaviour of the linear pair velocity dispersion. 
\end{abstract}
\section{Introduction.}
 The equations governing the evolution of the
statistical properties of disturbances in a critical ($\Omega=1$),
matter dominated universe are known to admit self-similar solutions
(Peebles 1980). This is because the universe expands as a power law of
time i.e. $a(t) \propto t^{\frac{2}{3}}$ and gravity itself does not
introduce any preferred scale. As a consequence of this it is
possible, using the linear theory of density perturbations, to
construct correlation functions which have a self-similar behaviour
over a range of scales. This method is valid only as long as the
correlations on these scales are extremely small. On the other hand it
is possible to use the stable clustering assumption to construct
correlation functions that have a self-similar behaviour over a range
of scales. The latter assumption is valid only on small scales where
virialized objects have already formed and clustering in real space
has ceased to increase. On scales where neither of these assumptions
can be used very little is known.

Hamilton et. al. (1991) have suggested a universal scaling relation
for the two point correlation function based on numerical evidence
from N-body simulations. In more recent papers Nityananda \& Padmanabhan
(1994), Peacock \& Dodds (1994), Jain et. al. (1995), Lokas et.al. (1995)
 and Padmanabhan (1996) have
investigated the proposed universal scaling relations. In an
earlier paper (Bharadwaj 1996) we have 
studied the lowest order non-linear corrections to the two point
correlation function for cases where the linear correlation function has a
self-similar behaviour over a range of scales. We restricted
ourselves to cases where the linear power spectrum has the form $P(k)
\propto k^n$ with $n\ge 0$ at small $k$ and we found that the non-linear
corrections do not have the same self-similar behaviour as the linear two
point correlation function. In a recent paper Scoccimarro and Frieman
(1996) have addressed the same question for initial conditions which
include $n<0$. For $n\ge 0$ their conclusions are similar to ours, but
they find that for 
$n<-1$ the lowest order  non-linear correction has the same
self-similar behaviour as the linear power spectrum. 

  In this paper we consider initial perturbations which
are a random Gaussian field and have a two point correlation function
which is self-similar over a range of scales. We study the
evolution of the correlation function analytically using the
Zel'dovich approximation (ZA) (Zel'dovich 1970) and we investigate under
what general conditions 
the non-linear corrections to the two point correlation function  exhibit
the same self-similar behaviour as 
the linear two point correlation function. ZA is known to be a very
good approximation to the full gravitational dynamics in the weakly
non-linear regime before the effects of multi-streaming become
important. It is thus expected that the results of this investigation
should also hold for gravitational dynamics and we have compared our
results with the results available from perturbative non-linear
gravitational dynamics (GD). The use of ZA instead of GD
makes the algebra much more tractable and it is
much simpler to interpret the results. It is also hoped
that the results of this investigation may help in building up
correlation functions that are self-similar in the non-linear epoch
starting from correlations that are self-similar in the linear regime.

The evolution of the various correlation functions in ZA has been studied
by earlier authors including Bond \& Couchman (1988), Grinstein \& Wise
(1987), and Schneider \& Bartlemann(1995 ). In  this paper we use the
notation  and some of the results of our earlier paper (Bharadwaj
1995) where we 
also presented some of our preliminary results regarding the non-linear
corrections to the two point correlation function in ZA and a comparison
with the results from GD.  

 \section{The two point correlation in ZA.}
 For a $\Omega=1$ universe ZA defines a map
 \begin{equation}
 x_{\mu}(t)=x_{\mu}(t_0) + a(t) u_{\mu} 
 \end{equation}
 from the initial position $x_{\mu}(t_0)$ of a particle to its position
$x_{\mu}(t)$ at some later  time $t$ in a comoving coordinate system. Here
the subscript $\mu$ takes values $1,2$ and $3$ corresponding to the
three Cartesian components and the Einstein summation convention holds
for it. The quantity $u_{\mu}$ is related to the peculiar velocity of
the particle and we shall refer to it as the velocity.

We are interested in the evolution of the statistical properties of
an ensemble of systems whose evolution is governed by ZA. In all the
members of the ensemble the particles are all initially uniformly
distributed and the initial velocity field is assumed to be
irrotational. It 
is also assumed that the velocity field in any member of the ensemble is a
particular realization of a Gaussian random field. These initial
conditions can be fully specified by the velocity-velocity correlation
which can be written in terms of a potential $\phi(x)$ as  
\begin{equation}
 <u_{\mu}(x^1)u_{\nu}(x^2)>= -\dl_{\mu} \dl_{\nu} \px
\end{equation} 
where $\dl_{\mu}=\frac{\dl}{\dl x_{\mu}}$ and 
\be x=\mid \vec{x}^1 -\vec{x}^2 \mid \,.  \ee

 For such an ensemble the two point correlation function can be
 written in a perturbative expansion as (Bharadwaj 1995)
\bea
 \xi(x,t)&=&\Sigma_{n=1}^{\infty} a^{2n} \frac{1}{n!} \dl_{\mu_1}
\dl_{\nu_1} \dl_{\mu_2} \dl_{\nu_2} ... \dl_{\mu_n} \dl_{\nu_n} \left[
\left( \dl_{\mu_1} \dl_{\nu_1} \px -\frac{1}{3} \delta_{\mu_1 \nu_1} \np0
\right) \right. \nonumber \\ & & \left.
 \left( \dl_{\mu_2} \dl_{\nu_2} \px -\frac{1}{3} \delta_{\mu_2
\nu_2} \np0 \right)...  \left( \dl_{\mu_n} \dl_{\nu_n} \px -\frac{1}{3}
\delta_{\mu_n \nu_n} \np0 \right) \right] \,. 
\label{eq:a1}
 \eea

The quantity that appears in the right hand side of this equation is
the dispersion of the pair velocity (i.e. the difference between the
velocity at the point $x^2$ and the point $x^1$)  which is defined as 
\bea
<v_{\mu} v_{\nu}>(x)&=& < (u_{\mu}(x^2)-u_{\mu}(x^1))( u_{\mu}(x^2)-
u_{\mu}(x^1)) > \nonumber \\
 &=& 2 \left( \dl_{\mu_2} \dl_{\nu_2} \px-\frac{1}{3} \delta_{\mu_2 \nu_2}
\nabla^2 \phi(0) \right) \,. \label{eq:a2}
\eea 
This quantity which we shall refer to the as the pair velocity
dispersion is related to the dispersion of the relative peculiar
velocities $\sigma_{\mu \nu}(x)$ in the linear epoch, and we have 
\be
\sigma_{\mu \nu}(x,t)=\lb a(t) \frac{d a(t)}{d t} \rb ^2<v_{\mu}
v_{\nu}>(x)  
\ee

 This dispersion arises due to the spread in the relative velocities
across the various realizations in the ensemble. The pair velocity
dispersion  is a symmetric tensor
and because the initial conditions are statistically homogeneous and
isotropic,  it can in general be written as
\be
<v_{\mu} v_{\nu}>(x)>= \de P(x) + \xmn Q(x) \label{eq:a3}
\ee
 where $P(x)$ is the dispersion of the relative velocity component
perpendicular to the separation $\vec{x}(=\vec{x}^2 -\vec{x}^1)$ and
$P(x)+Q(x)$ is the dispersion in the velocity component parallel to
$\vec{x}$. 

The two point correlation function can be written in terms of the pair
velocity dispersion as   
\bea 
\xi(x,t)=\Sigma_{n=1}^{\infty} \frac{1}{n!} 
\lb \frac{a^2}{2} \rb^n  & &\dl_{\mu_1} \dl_{\nu_1}
\dl_{\mu_2} \dl_{\nu_2} ... \dl_{\mu_n} \dl_{\nu_n} \left[ <v_{\mu_1}
v_{\nu_1}>(x) <v_{\mu_2} v_{\nu_2}>(x)  \right.  \nonumber \\ 
& & \left. ...  <v_{\mu_n}v_{\nu_n}>(x) \right] \,. \label{eq:a4}
\eea 

From this we obtain the linear two point correlation function as 
\be
\xiot=a^2 \xio= \frac{a^2}{2} \dl_{\mu_1} \dl_{\nu_1}
<v_{\mu_1}v_{\nu_1}> (x)  = a^2 \nf \px \,. \label{eq:a5} 
\ee

Given the linear two point correlation function we can invert equation
(\ref{eq:a5}) and express the initial pair velocity dispersion in
terms of the linear two point correlation function as
\bea
<v_{\mu} v_{\nu}>(x) = \frac{2}{3}\de \iox y dy \nonumber \\
-  \dl_{\mu} \dl_{\nu}(x) \iox y^2 dy  \nonumber \\
- \frac{1}{3} \dl_{\mu} \dl_{\nu}(1/x) \iox y^4 dy  \label{eq:a6}
\eea
and finally we are in a position to start investigating the nature of the
non-linear corrections to the two point correlation function, given the
linear two point correlation function. 

We consider situations where the linear two point correlation function at
some large scales i.e. for $x > x_1$ has the form 
\be
\xiot=a^2(t) \left( \frac{x_0}{x} \right)^{\gamma}=
\left( \frac{x_0(t)}{x} \right)^{\gamma}
\ee
where
\be
x_0(t) =x_0 a(t)^{\frac{2}{\gamma}}=x_0 \left( \frac{t}{t_0}
\right)^\frac{4}  {3 \gamma}  \,.
\ee 
We see that the two point correlation function has a self-similar
behaviour over a range of scales in the linear epoch and 
the effect of temporal evolution is to just scale the length scale
$x_0(t)$ that appears in the two point correlation function. 
 We want to investigate under what conditions the non-linear
corrections to the two point correlation function  has the same
self-similar behaviour.  Since the non-linear corrections are
determined by the pair velocity dispersion,  we proceed by first
investigating  the behaviour of the pair velocity dispersion. 
 
\section{The pair velocity dispersion.}
In this section  we investigate the spatial behaviour of the pair
velocity dispersion for different values of the index $\ga$. 
We separately consider the three different moments of the linear two point
correlation function that appear in equation (\ref{eq:a6}) for the
dispersion of the pair velocity  
\bea
A(x)= \iox y dy \\
B(x)= \frac{1}{x}\iox y^2 dy \\
C(x)=\frac{1}{x^3} \iox y^4 dy 
\eea
 It should be noted that all three of the
functions defined above have dimension  $L^2$ and it is possible that any
of them may introduce a new length scale in the evolution. 

\subsection{The behaviour of A.}
For $\gamma > 2$ the integral in A converges in the limit $x \rightarrow
\infty$ and at large $x$ we have
\be
A(x)= \ioi y dy  +\lb \frac{\x0g}{2- \gamma}\rb \frac{1}{x^{\gamma-2}} \,.
\ee
Here the first term is  a constant and the second term decays
as $x$ increases, and at large $x$ $A(x)$ tends to a constant value..

At the value $\gamma=2$ we have 
\be
A(x)= \ioo y dy  + x_0^2 \ln\left( \frac{x}{x_1} \right)
\ee
For $\gamma < 2$ we have 
\be
A(x)= \ioo y dy  - \left( \frac{\x0g}{2-\gamma}\right) x_1^{2-\gamma} +
\left( \frac{\x0g}{2-\gamma}\right)  x^{2-\gamma}  
\ee
where the first two terms are constants and the last term increases
monotonically with $x$ and  for large $x$ we have
\be
A(x)=\left( \frac{\x0g}{2-\gamma} \right) x^{2-\gamma}  
\ee

\subsection{The behaviour of B}
For $\gamma > 3$ the integral in B  converges as $x \rightarrow \infty$ and
 at large $x$ we have 
\be
B(x)= \frac{1}{x} \ioi y^2 dy + \lb \frac{\x0g}{3- \gamma} \rb
\frac{1}{x^{\gamma-2}} \,.
\ee
Here the first term falls as $\frac{1}{x}$ and the second term falls off
faster than this, and hence at large $x$ the first term dominates.

At the value $\gamma=3$ we have 
\be
B(x)=\frac{1}{x}\left[  \ioo y dy  + x_0^3 \ln\left( \frac{x}{x_1} \right)
\right]
\ee
and for large $x$ we have 
\be
B(x)=\frac{x_0^3}{x} \log\left( \frac{x}{x_1} \right)
\ee

and for $\gamma < 3$ we have 
\be
B(x)= \frac{1}{x} \left( \ioo y^2 dy  - \lb \frac{\x0g}{3-\gamma} \rb
x_1^{3-\gamma} \right) + \lb \frac{\x0g}{3-\gamma} \rb x^{2-\gamma}  
\ee
where for large $x$ the first terms falls off as $\frac{1}{x}$  and the
second term falls off slower. Thus at large $x$ we have  
\be
B(x)= \lb \frac{\x0g}{3-\gamma} \rb x^{2-\gamma}\,.  
\ee
where this is a decaying function for $2< \gamma < 3$, it is a constant
for $\gamma=2$ and it is a monotonically increasing function for
$\gamma<2$. 

\subsection{The behaviour of C}
For $\gamma  >5$ the integral in C converges as $x \rightarrow \infty$ and
 at large $x$ we have 
\be
C(x)= \frac{1}{x^3} \ioi y^4 dy + \lb \frac{\x0g}{5- \gamma} \rb
\frac{1}{x^{\gamma-2}} \,.
\ee
here the first term falls as $\frac{1}{x^3}$ and the second term falls off
faster than this, and hence at large $x$ the first term dominates.

At the value $\gamma=5$ we have 
\be
C(x)=\frac{1}{x^3}\left[  \ioo y^4 dy  +  x_0^5 \ln\left( \frac{x}{x_1}
\right) \right]
\ee
and for large $x$ we have 
\be
C(x)=\frac{x_0^5}{x^3} \ln\left( \frac{x}{x_1} \right)
\ee

For $\gamma < 5$ we have 
\be
C(x)= \frac{1}{x^3} \left( \ioo y^4 dy  - \lb \frac{\x0g}{5-\gamma} \rb 
x_1^{5-\gamma}  \right) + \lb \frac{\x0g}{5-\gamma} \rb  x^{2-\gamma}  
\ee
where at large $x$ the first terms falls off as $\frac{1}{x^3}$ and the
second term dominates. Thus at large $x$ we have  
\be
C(x)=\lb \frac{\x0g}{5-\gamma} \rb x^{2-\gamma}\,.  
\ee
where this is a decaying function for $2< \gamma < 3$, it is a constant
for $\gamma=2$ and it is a monotonically increasing function for
$\gamma<2$. 

\subsection{The spatial behaviour of the dispersion of pair velocities.}
Putting together the various components calculated earlier we can now
write expressions for the dispersion of the pair velocity for different
values of the index $\gamma$.

For the cases where $\gamma < 2 $ the behaviour is very simple and we have
\bea
<v_{\mu} v_{\nu}>(x)= & & \de \lb \frac{2}{3(2-\gamma)}
-\frac{1}{(3-\gamma)} 
+ \frac{1}{3(5-\gamma)}\rb \x0g x^{2-\gamma} \nonumber \\
&+& \xmn  \lb \frac{1}{3-\gamma}
- \frac{1}{5-\gamma}\rb \x0g x^{2-\gamma} \label{eq:c1}
\eea
For this case we see that both the components of the  pair velocity
dispersion increase as $\propto x^{2- \gamma}$.

Next, for  $\gamma=2$ we get 
\be
<v_{\mu} v_{\nu}>(x)=\frac{2}{3} x_0^2 \left[ \de \ln\lb \frac{x}{x_1} \rb 
+ \xmn \right] \label{eq:c2} \,.
\ee
Here  both the components of the pair velocity dispersion increase as
$\ln(x)$ and are 
nearly equal as we go to large separations

For $2 < \gamma < 3 $ we have to add an extra term 
\be 
\de l^2 = \frac{2}{3} \de \Ca \label{eq:c3}
\ee
to equation (\ref{eq:c1}). This term dominates the behaviour of the pair
velocity dispersion at large $x$ as the
contribution from the terms in equation (\ref{eq:c1}) get smaller as $x$
increases and  both the components of the pair velocity dispersion 
tend to a constant value $l^2$.  Thus we see that we have one new
length scale now i.e. $l$ and this plays a very crucial in the later
discussion. 

For $\gamma=3$ we have 
\bea
<v_{\mu} v_{\nu}>(x) =\frac{2}{3} \de \Ca - \de \frac{x_0^3}{x} \left[
\frac{1}{2} + \ln \lb \frac{x}{x_1} \rb \right] \nonumber \\
 -\xmn \frac{x_0^3}{x} \left[ \frac{1}{2} - \ln \lb \frac{x}{x_1} \rb
\right] \,. \label{eq:c4}
\eea

For $3 < \gamma < 5$, in addition to the expression (\ref{eq:c3}), we have
to also add 
\be
\left[ \xmn -\de \right] \frac{1}{x} \Cb \label{eq:c5}
\ee
to equation (\ref{eq:c1}).

For $\gamma=5$ we have 
\bea
<v_{\mu} v_{\nu}>(x)= \frac{2}{3} \de \ioi y dy 
+ \left[ \xmn -\de \right] \frac{1}{x} \ioi y^2 dy \nonumber \\
+ \de  \frac{x_0^5}{3 x^3} \left[ \frac{5}{6}+ \ln\lb \frac{x}{x_1}\rb
\right]
-\xmn  \frac{x_0^5}{ x^3} \left[ \frac{1}{2}+ \ln\lb \frac{x}{x_1}\rb
\right] \label{eq:c6}
\eea
For $\gamma > 5$ , in addition to the two terms in expressions
(\ref{eq:c3}) and (\ref{eq:c5})  we have to also add 
\be 
\left[ \frac{1}{3} \de - \xmn \right] \frac{1}{x^3} \Cc
\ee
to equation (\ref{eq:c1}).

 Finally, we see that for $\gamma < 2$ the dispersion in the pair velocity
keeps on increasing as a power law. For $\gamma=2$  it increases
logarithmically with the separation and for all $\gamma \le 
2$ pair velocity dispersion  tends to infinity as the separation $x$ tends to
infinity. For  $\gamma > 2$ we find that the pair
velocity dispersion reaches  the constant value $l^2$ which is
determined by the two point correlation at small separations.   

We next consider some specific cases where the linear power spectrum
is of the form $P(k)= A e^{-k} k^n$ with $n=(a)\,-2,(b)\,-1,(c)\,0$ and
$(d)\,1$. The corresponding values of $\gamma$ are
$(a)\,1,(b)\,2,(c)\,4$ and 
$(d)\,4$. For the case with  $n=0$ the large $x$ behaviour is decided
by the exponential cutoff in $P(k)$ and because of this it has the same power
law index $(\gamma=4)$ as the $n=1$ case. 
Figure (1) shows the tangential component of the pair velocity
dispersion for these four cases and it illustrates the main point of this
section.  The behaviour of the radial component of the pair velocity
dispersion is very similar. 

\section{The non-linear corrections to $\xi$.}
Here we use the pair velocity dispersion calculated in the previous
section to  investigate the nature of the  non-linear corrections
to the two point correlation function. 
From equation (\ref{eq:a6})  we obtain the lowest order non-linear
correction to the two point correlation function as
\bea 
\xitt= a^4 \xit = 
\frac{a^4}{8} \dl_{\mu_1} \dl_{\nu_1}
\dl_{\mu_2} \dl_{\nu_2}  \left[ <v_{\mu_1} v_{\nu_1}>(x) 
<v_{\mu_2} v_{\nu_2}>(x)  \right]  \label{eq:d1} \,.
\eea 

For $\gamma<2$ the lowest order non-linear correction to the two point 
correlation function is 
\be
\xit=\frac{2(285-679 \ga + 611 \ga^2 - 259 \ga^3 +52 \ga^4 -4 \ga^5)} { (2-
\ga) (3- \ga)^2 (5- \ga)^2 } \lb \frac{x_0}{x} \rb ^{2 \ga} \,.
\label{eq:d2} 
\ee
and  for this range of $\gamma$ we find that $\xitt \propto \left[
\xiot \right] ^2 $.   For $\gamma<2$  the pair
velocity dispersion is a power law in $x$  and it is $\sim x^{2-\gamma}$. To
calculate the $n^{th}$ order non-linear
corrections to the two point correlation function we take $n$ of these
functions  i.e. an expression  $\propto (x^{2-\gamma} )^n $  and we act on
this with $2n$ spatial derivatives and the result is $x^{-n \gamma} 
\propto \left[ \xio \right]^n$. Thus we see that $n$th order non-linear
correction is of the form  $\xi^{(n)}(x) \propto
[\xi^{(1)}(x)]^n$  and we can write the two point correlation function in a
series of the form
\be
\xi(x,t)= \sum_{1}^{\infty} c_n(\ga) \left[ \xiot \right]^n 
\ee
where the $c_n(\ga)$ are coefficients which depend only on $\gamma$.
We see that for  $\gamma<2$ the  non-linear corrections to
the two point correlation function have the same self-similar behaviour as
the linear two point correlation function.  

For $\ga=2$ we have 
\be
\xit=\frac{2}{3} \lb \frac{x_o}{x} \rb^4 \ln \lb \frac{x}{x_1} \rb \,.
\label{eq:d3}
\ee
For this case we find that the non-linear correction cannot be
written in terms of the linear two point correlation alone and it
does not have a self-similar behaviour.

 These are the only cases where the pair velocity dispersion keeps on
increasing at large $x$. For larger values of $\gamma$ the pair velocity
dispersion reaches a constant value and as consequence the behaviour of
the two point correlation too is different.  For all higher values of
$\ga$, in addition to  the contribution to $\xit$ given by  equation
(\ref{eq:d2}) which behaves as $x^{(4-2 \gamma)}$,  there are other terms
which fall of slower and dominate the behaviour at large $x$. 
For $(2< \gamma <3)$ we have
\be
\xit=\left[ \frac{1}{3} \Ca \right] \ga (\ga-1) \frac{x_0^{\ga}}
{x^{2+\gamma}} \,. \label{eq:d4} 
\ee
For these cases we see that the self-similar behaviour is broken.
This is because the pair velocity dispersion reaches a
constant value and this introduces a new length scale $l$ whose square
appears in the square brackets in equation (\ref{eq:d4}). This length
scale $l$  changes the scaling property of the two point correlation
function.  

For $\gamma>3$ we have to take into account one more term and we have 
\be
\xitt=\left[ \frac{1}{3} \Ca\right] \ga (\ga-1)
\frac{x_0^{\ga}}{x^{2+\gamma}} 
+\frac{3}{x^6} \left[  \Cb \right] ^2 \,. \label{eq:d5}
\ee

For $\gamma <4$ it si obvious that the first term falls off slower than
$x^{-6}$ and it dominates the large $x$ behaviour. For $\gamma=4$ both the
terms have a $x^{-6}$ behaviour and we cannot  drop any one of them. For
$\gamma >4$ the first term falls off faster than $x^{-6}$ and we would
expect the second term to dominate, but for many of the cases of interest
the integral $\ioi y^2 dy$ is zero and then it is only the first term that
contributes. This is illustrated by the two cases $(c)$ and $(d)$ that we
have considered earlier. The moments of the two point correlation function
can be written in terms of the power spectrum as 
\be
\ioi y dy=\frac{1}{2 \pi^2} \int_0^{\infty} P(k) dk
\ee
 and
\be
\ioi y^2 dy=\frac{P(0)}{4 \pi} \,.
\ee
For both $(c)$ and $(d)$ we have $\gamma$=4, but using the above equations
we find that  for $(c)\, (i.e. n=0)$ we obtain $\ioi y^2 dy= \frac{A}{4 \pi}$
and for $(d)\, (i.e. n=1)$ we have  $\ioi y^2 dy= 0$, and for both of them
we have  $\ioi y dy= \frac{A}{2 \pi^2}$. Thus  for $(c)$ the second term
in equation (\ref{eq:d5}) contributes and for $(d)$ it is zero and for
these two cases we obtain 
\be
\rm{for}\, (a)\hspace{1.in} \xit=\frac{A}{\pi^2} \lb 2 x_0^4 + \frac{3
A }{16} \rb \frac{1}{x^6} 
\ee
and
\be
\rm{for} \,(b) \hspace{1.in} \xit= \frac{2 A x_0^4}{\pi^2}   \frac{1}{x^6}
\,.
\ee
Finally we can generalize this to say that for  all cases where linear
power spectrum is of the form $P(k) \propto k^n$ with $n>-1$ for small $k$,  
the lowest order non-linear correction to the two point correlation
function is given by equation  (\ref{eq:d4}), except for the case when
$n=0$. For $n=0$ we have to take into account the extra term in equation 
(\ref{eq:d5}).

	In a recent paper Taylor and Hamilton (1996) have considered the 
evolution of the non-linear power spectrum in the Zel'dovich approximation. 
They present exact analytic results for cases where the linear power spectrum
is a power law with the power law index $n=-2,-1$ i.e. $(\gamma=1,2)$.
We have compared 
our result at the lowest order of non-linearity with the corresponding result 
implied by the exact expression calculated by Taylor and Hamilton (1996). 
We find that while the two match for $n=-2$, there is a disagreement for
$n=-1$. This is because for $n=-1$ $(\gamma=2)$ they have considered the pair
velocity dispersion to bee a constant value whereas we find that it has a
logarithmic dependence on the separation.    
In addition to this, they have considered $n=-1$ as a limiting case of a 
situation where the power law index is of the form $n=-1-\epsilon$. 
In our study here we find that the behaviour of the case with $\gamma=2$ i.e. 
$n=-1$ is quite different from the results for  $\gamma <2 $ and we do not 
expect that the limit taken by Taylor and Hamilton (1996) will give the 
correct result for  $n=-1$.

\section{Discussion and Conclusion.}
We have considered linear two point correlation functions that have the
form $\frac{1}{x^{\gamma}}$ at large separations. For $\Omega=1$ they have
a self-similar behaviour in the linear epoch. We have investigated under
what conditions the non-linear corrections calculated using the Zel'dovich
approximation have the same self-similar behaviour.
     
We find that the scaling properties of the non-linear corrections to the
two  point correlation function are
determined by the spatial behaviour of the linear pair velocity dispersion. 
For $\gamma <2$ both the radial and tangential components of the pair
velocity dispersion keep on increasing as
$x^{2-\ga}$ and it has a local dependence on the linear two point correlation
function i.e. the linear pair velocity dispersion at some separation $x$
depends only only the linear two point correlation at the same separation.
 As a consequence all the non-linear corrections to the two point
correlation function also have a local dependence on the linear two point
correlation function and the $n$ th order non-linear
correction to the correlation function  $\xi^{(n)}(x,t)$ has the property 
 $\xi^{(n)}(x,t) \sim [\xiot]^n$. We see that for these cases all the
non-linear corrections have the same self-similar behaviour as 
linear  two point correlation.

For $\ga=2$ the pair velocity 
dispersion increases logarithmically 
at large $x$ and the $n$ th order non-linear correction is of the form 
$\frac{[\ln(x)]^{n-1}}{x^{n \ga}}$ i.e. $\xi^{(n)}(x,t) \sim [\ln\lb \frac
{x}{x_1} \rb ]^{n-1} [\xiot]^n$. Because of the  extra $[\ln \lb
\frac{x}{x_1} \rb]^{n-1}$ the  non-linear 
corrections do not have the same self-similar behaviour as the linear two
point correlation function. 

 For $\gamma>2$ the pair velocity dispersion
at large $x$ reaches a constant  
value $l^2$ which is determined by the linear two point correlation
at small scales. For this case the linear pair velocity dispersion has a
non-local dependence on the linear two point correlation function. This
also introduces a new length scale and the  $n$ th order correction is of
the form  $\frac{l^{2(n-1)}}{x^{\ga+2(n-1)}}$. This has completely
different scaling properties and the non-linear corrections do not have
the same self-similar behaviour as the linear two pint correlation.

 In an earlier paper (Bharadwaj 1996) we have studied the non-linear
correction to the two point correlation function using perturbative
gravitational dynamics. We only considered cases  with $\ga >2$
and we found that the non-linear corrections do not have the same
self-similar behaviour as the linear two point correlation function. We
also found that this was due to the emergence of a new length scale $l$
which appears here also and we interpreted this in terms of a simple
diffusion process. Scoccimarro \& Frieman (1996) have studied the lowest
order non-linear correction to the power spectrum  and the  corresponding
correlation functions include those with $\gamma<2$. They find that for
power spectra where the index $n$ is less than $-1$ (or $\gamma <2$), the
lowest order non-linear correction has the same self-similar behaviour as
the linear power spectrum and  they also find that the self-similarity is
broken for $n\ge -1$ (i.e.$ \gamma \ge 2$).    We see that  ZA makes the
same predictions as  GD regarding the scaling properties of the
non-linear corrections to the two point correlation function and we expect
that any conclusions that can be drawn on the basis of our investigations
using ZA should also hold for GD. We find that in ZA the scaling
properties of the non-linear corrections are decided by the spatial
behaviour of the linear 
pair velocity dispersion. We propose that this is also true for the
non-linear corrections calculated using perturbative gravitational
dynamics. Although the relation between the non-linear corrections and the
linear pair velocity dispersion is quite explicit in the Zel'dovich
approximation (equation \ref{eq:a4}), it is not clear how this comes about
in perturbative gravitational dynamics.

The pair velocity dispersion is the lowest velocity moment of the two
point distribution function that has information about the velocity
components tangential to the separation $\vec{x}$. It is very interesting
that it is this quantity, and not some lower velocity moment, that turns
out to play an important role in deciding the scaling properties of the
non-linear corrections.
\acknowledgements The author would like to thank Rajaram
Nityananda for many useful discussions and T.R. Seshadri for his
encouragement. 

\newpage
\begin{figure}
\plotone{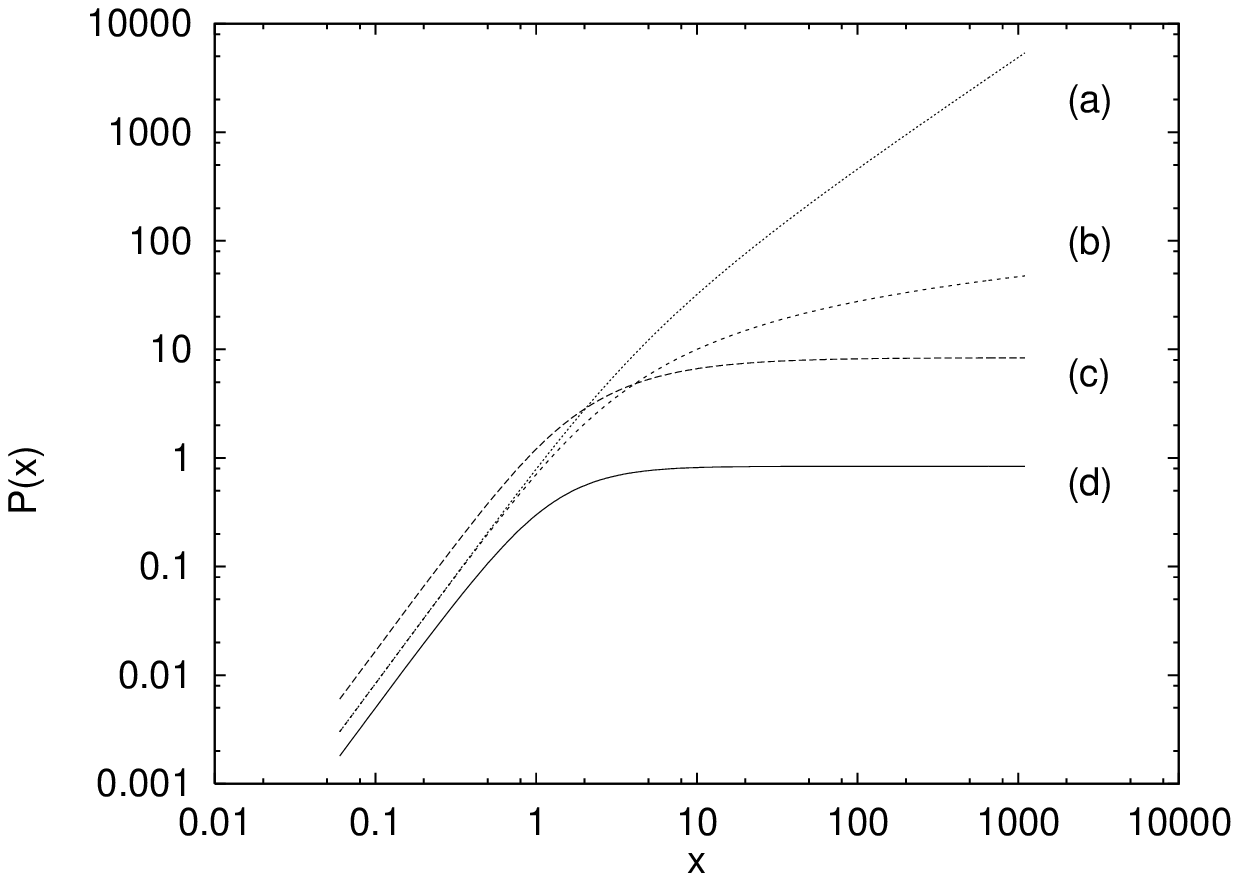}
\caption{The tangential component of the pair velocity dispersion P(x)
shown as a function of the separation $x$ for the four cases discussed in
section 3.4.}
\end{figure}
\end{document}